\documentclass[aps,prl,twocolumn,longbibliography,superscriptaddress,10pt]{revtex4-1}

\usepackage{amsfonts,amssymb,dsfont}
\usepackage[sumlimits,intlimits]{amsmath}
\usepackage{graphics}
\usepackage{graphicx}
\usepackage{color}
\usepackage{mathrsfs}
\usepackage{textcomp}
\usepackage{verbatim}
\usepackage[colorlinks=true,citecolor=blue]{hyperref}
\usepackage{braket}
\usepackage{times}
\usepackage[T1]{fontenc}
\usepackage[final]{pdfpages}
\usepackage{pgffor}
\makeatletter
\AtBeginDocument{\let\LS@rot\@undefined}
\makeatother

\usepackage[capitalise,compress]{cleveref}
\crefrangelabelformat{equation}{\textup{(#3#1#4)}--\textup{(#5#2#6)}}

\DeclareMathOperator{\tr}{tr}

\begin{document}

\title{Probing ground-state phase transitions through quench dynamics}

\author{Paraj Titum}
\affiliation{Joint Quantum Institute, NIST/University of Maryland, College Park, Maryland 20742, USA}
\affiliation{Joint Center for Quantum Information and Computer Science, NIST/University of Maryland, College Park, Maryland 20742, USA}

\author{Joseph T. Iosue}
\affiliation{Joint Quantum Institute, NIST/University of Maryland, College Park, Maryland 20742, USA}
\affiliation{Massachusetts Institute of Technology, Cambridge, Massachusetts 02139, USA}

\author{James R. Garrison}
\affiliation{Joint Quantum Institute, NIST/University of Maryland, College Park, Maryland 20742, USA}
\affiliation{Joint Center for Quantum Information and Computer Science, NIST/University of Maryland, College Park, Maryland 20742, USA}

\author{Alexey V. Gorshkov}
\affiliation{Joint Quantum Institute, NIST/University of Maryland, College Park, Maryland 20742, USA}
\affiliation{Joint Center for Quantum Information and Computer Science, NIST/University of Maryland, College Park, Maryland 20742, USA}

\author{Zhe-Xuan Gong}
\affiliation{Department of Physics, Colorado School of Mines, Golden, Colorado 80401, USA}
\affiliation{Joint Quantum Institute, NIST/University of Maryland, College Park, Maryland 20742, USA}
\affiliation{Joint Center for Quantum Information and Computer Science, NIST/University of Maryland, College Park, Maryland 20742, USA}

\date{\today}

\begin{abstract}
The study of quantum phase transitions requires the preparation of a many-body system near its ground state, a challenging task for many experimental systems. The measurement of quench dynamics, on the other hand, is now a routine practice in most cold atom platforms. Here we show that quintessential ingredients of quantum phase transitions can be probed directly with quench dynamics in integrable and nearly integrable systems. As a paradigmatic example, we study global quench dynamics in a transverse-field Ising model with either short-range or long-range interactions. When the model is integrable, we discover a new dynamical critical point with a non-analytic signature in the \textit{short-range correlators}. The location of the dynamical critical point matches that of the quantum critical point and can be identified using a finite-time scaling method. We extend this scaling picture to systems near integrability and demonstrate the continued existence of a dynamical critical point detectable at prethermal time scales. Therefore, our method can be used to approximately locate the quantum critical point.  The scaling method is also relevant to experiments with finite time and system size, and our predictions are testable in near-term experiments with trapped ions and Rydberg atoms.
\end{abstract}

\maketitle

{\it Introduction.}---%
Experimental advances in isolating and controlling nonequilibrium quantum systems have brought  within reach answers to many fundamental questions, including those related to thermalization, prethermalization, and many-body localization \cite{Deutsch1991-eth, Srednicki1994-thermalization, Srednicki1999-eth, Rigol2008-eth, Kim2014-all-eigenstates, Garrison2018-eth}. The exquisite control of complex quantum systems has become commonplace as a result of progress in various platforms, such as trapped ions \cite{richerme2014, neyenhuis2017}, ultracold  atoms \cite{kinoshita2006,schreiber2015}, nitrogen-vacancy centers~\cite{Lukin_NV_2017}, Rydberg atoms \cite{Lukin_Rydberg_2017}, and others. 

Among other interesting topics in nonequilibrium quantum many-body physics,  phase transitions that emerge in the dynamics of isolated quantum systems have attracted significant theoretical and experimental interest \cite{Heyl_Plkovnikov_Kehrein_DPT_2013, zunkovic2018,heyl2018,Jurcevic_DPT_expt_2017,Zhang_Monroe_2017,halimeh2017,halimeh2017a,homrighausen2017}. There are two known types of dynamical phase transitions: (i) when a global order parameter (such as the Loschmidt echo) shows an abrupt change as a function of evolution time, or (ii) when a local order parameter measured after a sufficiently long time becomes nonanalytic as a function of some Hamiltonian parameter \cite{zunkovic2018}. This latter type of dynamical phase transition is closely related to quantum phase transitions; the only difference is that the order parameter is measured in the quenched state instead of the ground state. It is thus natural to ask how this dynamical phase transition is related to a quantum phase transition. While difficult to answer, this question is not only of fundamental importance, but also motivates the idea of using dynamics to study quantum phase transitions. In fact, in many of the above-mentioned experimental platforms, cooling a system to its ground state can be a formidable task while performing a quench experiment is now a routine practice.

In this Letter, we first establish a strong connection between the quantum critical point and a new dynamical critical point in a general class of integrable models, using the transverse-field Ising model (TFIM) as a paradigm. We show analytically that these critical points are identical and expect such behavior to generalize to other systems consisting of noninteracting particles~\cite{Roy_Moessner_Das_2017}. This dynamical critical point has a nonanalytic signature in the long-time values of short-range, two-point correlation functions. Much of the previous work on the quench dynamics of the TFIM has focused on the behavior of long-range correlations~\cite{Calabrese_Essler_Fagotti_PRL_2011,Calabrese_Essler_Fagotti_JStat-1,Calabrese_Essler_Fagotti_JStat-2,zunkovic2018,halimeh2017,Karl_Kastner_quench_2017}, a common practice in studying equilibrium phase transitions. However, these correlations vanish in the thermodynamic limit for all nonzero field values due to the absence of long-range order at long time. Thus short-range correlations, often ignored due to the dominance of long-wavelength physics at low temperature, are important in identifying dynamical criticality and revealing its connections to quantum phase transitions.

Second, we show that analogous to the finite-size scaling analysis for identifying quantum critical points, one can perform a finite-time scaling analysis for obtaining the dynamical critical point. Intuitively, this can be understood because the evolution time controls the effective system size seen by short-range correlations as a result of emergent light cones. This finite-time scaling analysis is particularly favorable for quantum simulation experiments, as one does not need to create systems of different sizes or wait for a time much longer than the coherence time, and thus, allows for near-term experimental demonstration \cite{Zhang_Monroe_2017,Lukin_Rydberg_2017}.

Finally, we generalize our findings to systems that are nonintegrable by adding weak interactions. We show that the finite-time scaling predicts a dynamical critical point in the prethermal time scale. In general, the dynamical critical point will no longer coincide with the quantum critical point away from integrability. But, as shown by our analysis of the TFIM with next-nearest-neighbor and power-law decaying interactions, we expect the two transition points to be close to each other when interactions are weak. As perturbation theory may not work for finding quantum critical points of weakly interacting systems, our findings provide an alternative and experimental way of locating such critical points. 

We point out that some earlier works have considered similar ideas of studying quantum criticality via quench dynamics. For example, Ref.~\cite{Prosen_2000} studies the appearance of nonanalytic signatures in the periodically kicked Ising model, Ref.~\cite{Roy_Moessner_Das_2017} discusses nonanalytic signatures in the long-time dynamics in noninteracting topological phase transitions, and Ref.~\cite{Heyl_OTOC_2018} uses out-of-time-order correlators (OTOCs) to identify quantum phase transitions. However, our approach offers three unique advantages: (i) The short-range correlations are easy to measure experimentally, especially compared to the OTOC\@. (ii) Our approach is not restricted to exactly integrable systems. (iii) The finite-time scaling analysis we introduce provides a method to locate the dynamical critical point.

{\it Model.}---
We consider two models for the quench Hamiltonian of $L$ spins in one dimension: a transverse-field Ising model with either next-nearest-neighbor or long-range interactions,
\begin{align}
H_\text{NNN}&=-J\sum_{\langle i,j\rangle}\sigma_{i}^{x}\sigma_{j}^{x}-J\Delta \sum_{\langle\langle i,j\rangle\rangle}\sigma_{i}^{x}\sigma_{j}^{x}+B\sum_{i}\sigma_{i}^{z},\label{eq:model-NNN}\\
H_\text{LR}&=-\sum_{i<j} J^{(\alpha)}_{ij}\sigma_{i}^{x}\sigma_{j}^{x}+B\sum_{i}\sigma_{i}^{z},
\label{eq:model-LR}
\end{align}
where $\{\sigma_i^{x,y,z}\}$ denote the Pauli matrices and $\langle\cdots\rangle$ and $\langle\langle\cdots\rangle\rangle$ denote nearest and next-nearest neighbors, respectively. We will use periodic boundary conditions to ensure translation invariance unless otherwise noted. In the long-range Hamiltonian, the Ising coupling is defined as $J^{(\alpha)}_{ij}=J\left[\frac{1}{\left|i-j\right|^{\alpha}}+\frac{1}{\left|L-(i-j)\right|^{\alpha}}\right]$, which accounts for  periodic boundary conditions~\cite{Sandvik2003}. We restrict to case of ferromagnetic interactions with $J$ and $\Delta>0$.

\begin{figure}
\begin{center}
\includegraphics[width=\linewidth]{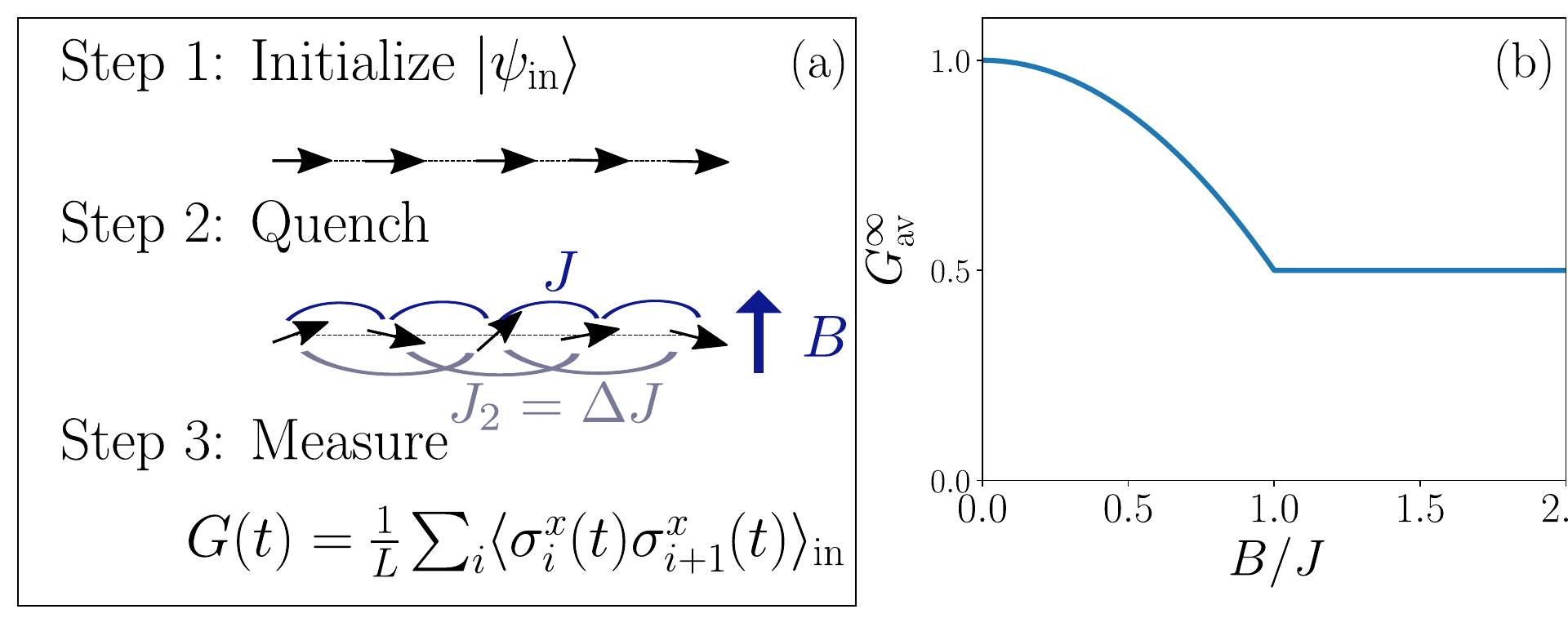}
\end{center}
\caption{(a) Schematic picture of the quench. We evolve the initial state $|\psi_{\rm in}\rangle$, which consists of all spins polarized in the Ising direction, for a time $t$ under the Hamiltonian defined in \cref{eq:model-NNN} (shown) or \cref{eq:model-LR}. Then we measure the nearest-neighbor correlator  $G(t)$ defined in \cref{eq:NNCorr-def}. (b) The dependence of $G^\infty_{\rm av}$ on $B$ as given by \cref{eq:G_GGE_formula}. }
\label{fig:model} 
\end{figure}

The quench and measurement protocol is shown schematically in \cref{fig:model}(a). We initialize the system in a product state with all spins polarized in the Ising direction, $|\psi_{\rm in}\rangle= \ket{\rightarrow \cdots \rightarrow} $. This state is one of two degenerate ground states when $B=0$. We focus on the dynamics under the quench Hamiltonian [see \cref{eq:model-NNN,eq:model-LR}] of the nearest-neighbor, equal-time correlation function defined as
\begin{equation}
G(t)=\frac{1}{L}\sum_{\langle i,j\rangle} \langle \sigma^x_i (t) \sigma^x_{j}(t)\rangle_{\rm in},
\label{eq:NNCorr-def}
\end{equation}
where $\langle\cdots \rangle_{\rm in}$ indicates the expectation value with respect to the initial state $|\psi_{\rm in}\rangle$ defined above, and the operators are written in the Heisenberg picture, $\sigma^x_i (t)=e^{iHt}\sigma^x_ie^{-iHt}$. We note that the dynamics of two-point correlators at longer distances independent of the system size (e.g.\ $\langle\sigma_i^x\sigma_{i+2}^x\rangle$) will be similar to $G(t)$. However, calculating such longer range correlations would require a more complicated analytical treatment. 

{\it Results at the integrable point.}---%
Consider first the nearest-neighbor TFIM [$\Delta=0$ in \cref{eq:model-NNN} or $\alpha=\infty$  in  \cref{eq:model-LR}]. In this case, the Hamiltonian is integrable and can be mapped to free fermions via a Jordan-Wigner transformation with the boundary condition determined by the particle number parity~\cite{supp}. The quasiparticle dispersion is given by $\omega_{q}=2\sqrt{B^{2}-2BJ\cos q+J^{2}}$, where $q$ denotes momentum, and the Hamiltonian becomes $H=\sum_q\omega_q \gamma^\dagger_q\gamma_q$, where $\gamma_q$ are the annihilation operators for the Bogoliubov quasiparticles. The time evolution is governed by a set of conserved densities of these quasiparticles, $I_q=\gamma^\dagger_q\gamma_q$. The ground state of this model exhibits a second-order phase transition with the critical point at $B=B_c^{\rm gs} \equiv J$.

We calculate the time-averaged correlator $G_{\text{av}}(t)\equiv\frac{1}{t}\int_0^t d\tilde{t}\,G(\tilde{t})$ to be~\cite{supp}
\begin{equation}
G_{\text{av}}(t)=\frac{1}{L}\sum_{q}\frac{8B^2}{\omega_{q}^{2}}\left[\left(\frac{J}{B}-\cos q\right)^{2}+ j_0\left(2\omega_{q}t\right)\sin^{2} q\right],
\label{eq:time-avgd-corr}
\end{equation}
where $ j_0(z)=\sin \left(z\right)/z$ is a spherical Bessel function of the first kind. Note that the second term in the correlator decays at long times: $j_0\left(2\omega_{q}t\right)\rightarrow 0$ as $t\rightarrow \infty$ (the only exception being the $\omega_q = 0$ case which contributes a time-independent constant $\propto 1/L$). This means that the first term in \cref{eq:time-avgd-corr} corresponds to the long-time stationary value. Taking the thermodynamic limit, we obtain an analytic expression for $G_{\rm av}(t)$,
\begin{align}
G^\infty_{\rm av}&\equiv\lim_{\begin{smallmatrix}
L\rightarrow\infty\\
t\rightarrow\infty
\end{smallmatrix}}\frac{1}{t}\int_0^t d\tilde{t}\,G(\tilde{t})=\begin{cases}
1-\frac{B^{2}}{2J^{2}} & \text{if }B\leq J\\
\frac{1}{2} & \text{if }B\geq J,
\end{cases} 
\label{eq:G_GGE_formula}
\end{align}
which has a nonanalyticity at $B=B_c^{\rm dy} \equiv J$. We refer to this nonanalytic point as a dynamical critical point. Note that the dynamical and ground state critical points are identical, $B_c^{\rm dy}=B_c^{\rm gs}$. This is because the nonanalyticity can be traced to the appearance of $1/\omega_q^2$ in the expression for $G_{\rm av}(t)$ [see \cref{eq:time-avgd-corr}], which has a pole at $B=B_c^{\rm gs}$.
In \cref{fig:model}(b), we plot the long-time value of the correlator $G^\infty_{\rm av}$ [\cref{eq:G_GGE_formula}], which exhibits a kink at $B=B_c^{\rm dy}$. Note that the expression for $G^\infty_{\rm av}$ is identical to that obtained from the Generalized Gibbs Ensemble (GGE)~\cite{Vidmar_Rigol_2016} for the TFIM. 

\begin{figure}
\includegraphics[width=\linewidth]{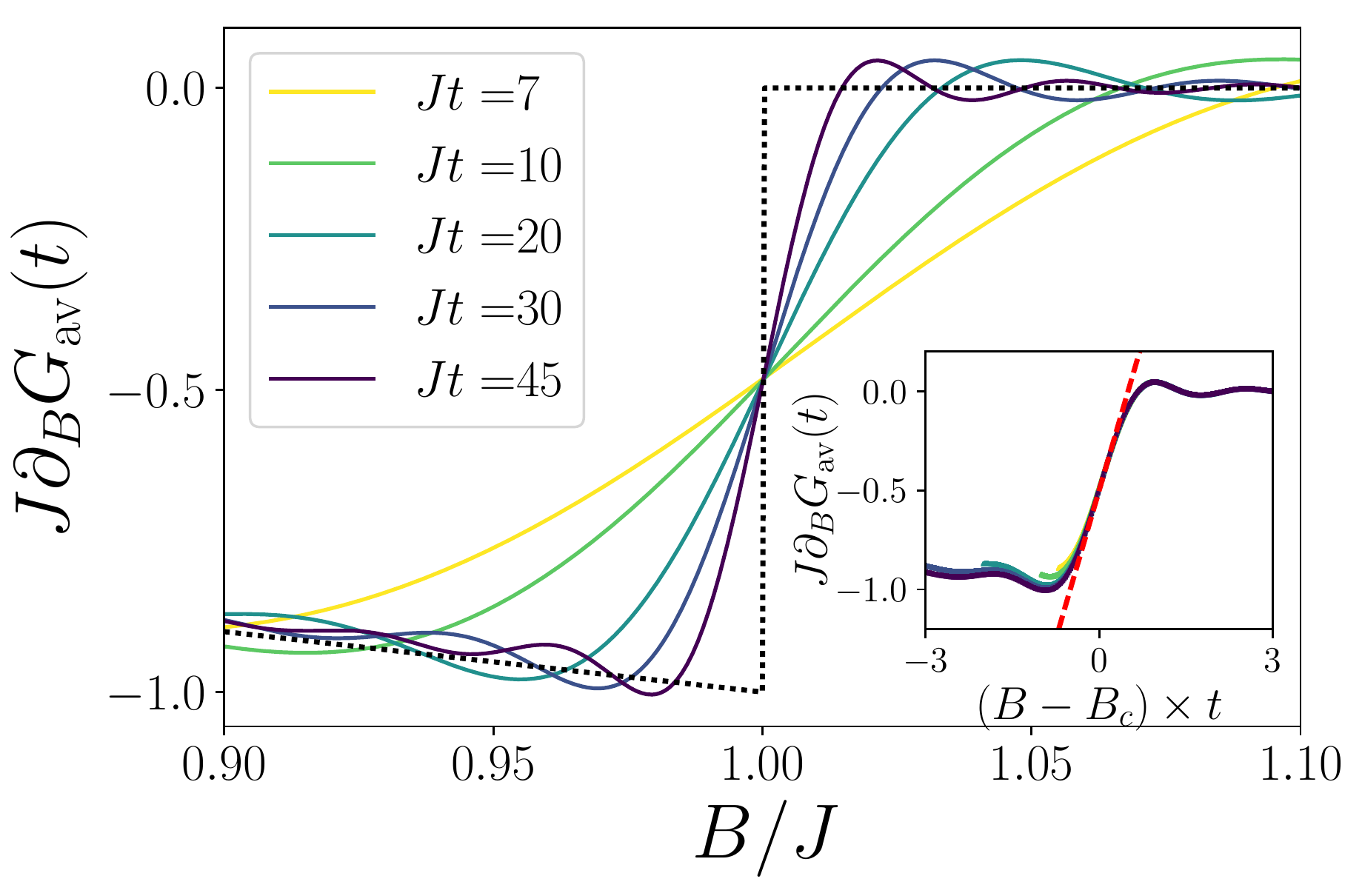}
\caption{Finite-time scaling for the integrable case. The first derivative of the correlator, $\frac{\partial}{\partial B}G_{\rm av}(t)$,  exhibits a sharper jump across the transition ($B=B_c^{\rm dy}=J$) at later times. The curves are obtained by differentiating \cref{eq:time-avgd-corr} with a system size $L=100$~\cite{supp}. The curves at different times cross at the same point, thus revealing a dynamical critical point. The dotted line in black indicates the expected discontinuity in the derivative in the thermodynamic limit ($L\rightarrow \infty$) and at long times ($Jt\rightarrow\infty$). (Inset) We extract the time dependence by performing a scaling collapse after rescaling the magnetic field $B$ by $t$. The scaling function near the critical point is linear, and is shown as a dashed red line.}
\label{fig:finite-time-scaling}
\end{figure}

The kink at the dynamical critical point is obtained after taking two limits in either order: (i) the infinite-time limit and (ii) the thermodynamic limit. In any realistic experiment, one can only measure $G_{\rm av}(t)$ at a finite time and a finite system size. In this case, $G_{\rm av}(t)$ is a smooth function of $B$, but we can nevertheless locate the dynamical critical point in the following way:  We obtain the time-dependent derivative of $G_{\rm av}(t)$ with respect to $B$, i.e.\ $\partial_B G_{\rm av}(t)$  (see \cite{supp} for the explicit expression), and plot it in \cref{fig:finite-time-scaling}. We find the curves at different times cross at the same point  up to a small correction $\propto\frac{1}{L}$, thus revealing a dynamical critical point. In fact, all the curves can be made to collapse into a single curve (shown in the inset) by rescaling the field $B$ by $Jt$. The expression for the first derivative near the dynamical critical point takes the form of a universal scaling function, $\partial_B G_{\text{av}}(t)=\frac{1}{J}f((B-B_c^{\rm dy})t)$. To lowest order in the distance from the critical point (given by $\epsilon=B-B_c^{\rm dy}$), the scaling function is $f(\epsilon t)=-\frac{1}{2}+\epsilon t$~\cite{supp}. Note that this finite-time scaling function  is very similar to the finite-size scaling function of the long-time correlator near the critical point, with $t$ playing the role of $L/(3J)$ \cite{supp}. The finite-time scaling analysis thus allows us to locate the position of the dynamical critical point, and in this (integrable) case, also the quantum critical point. 

Let us discuss the generality of this result for different initial states. For an arbitrary initial state, $G^\infty_{\rm av}$ is given by
\begin{equation}
G^\infty_{\rm av,arbitrary}=\frac{1}{L}\sum_{q}\frac{2(J-B\cos q)}{\omega_{q}}\left[1-2\left\langle I_{q}\right\rangle _{\rm in}\right],\label{eq:G-arbitrary-initial-state}
\end{equation}
where $\left\langle I_{q}\right\rangle_{\rm in}$ is the expectation value of the conserved quasiparticle densities in the initial state. It is clear from this expression that the nonanalyticity which results from a gapless $\omega_q$ will survive for arbitrary initial states unless the form of $\left[1-2\left\langle I_{q}\right\rangle _{\rm in}\right]$ cancels off the $1/\omega_q$ singularity at $q= 0$. A generic pure initial state, in particular a product state that can be easily prepared experimentally, will thus lead to the same dynamical critical behavior. Interestingly, a thermal initial state given by the density matrix $\rho_{\rm th}=e^{-\beta H}/\tr \left[e^{-\beta H}\right]$ is an exception. This is because in this case, $\langle I_q\rangle_{\rm in}=\tr (\rho_{\rm th} I_{q})=\frac{1}{2}\left[1-\tanh\left(\frac{\beta\omega_{q}}{2}\right)\right]\approx \frac{1}{2}\left(1-\frac{\beta}{2}\omega_q\right)$ when $\omega_q\rightarrow 0$ and thus $G^\infty_{\rm av}$ becomes an analytic function of $B$. This is consistent with the well-known fact that the 1D nearest-neighbor TFIM does not have a thermal phase transition. 

{\it Results away from integrability.}---%
Now, let us consider the TFIM with either an additional next-nearest-neighbor interaction  [$\Delta \neq 0$ in \cref{eq:model-NNN}] or long-range interactions [any finite $\alpha$ in \cref{eq:model-LR}]. Assuming the Eigenstate Thermalization Hypothesis (ETH)~\cite{Srednicki1999-eth} holds, we expect local observables, including $G(t)$, to thermalize at sufficiently long times. For any finite value of $\Delta$ or $\alpha>2$, it has been established that the TFIM does not exhibit a thermal phase transition \cite{sachdev_2011,Dutta_Bhattacharjee_2001}. Thus, once the system fully thermalizes, any measured observable has to be analytic and thus, it should no longer have a dynamical critical point. (The situation with $\alpha<2$ is beyond the scope of this Letter due to the strong effects of long-range interactions that could break ETH~\cite{zunkovic2018}.)

However, as recently shown by Refs.~\cite{halimeh2017,halimeh2017a}, it is possible to still observe a dynamical phase transition when local observables have not yet thermalized but instead have reached prethermal values. This is the case here when $\Delta$ is small or $\alpha$ is large, so that the Hamiltonian is nearly integrable and prethermalization is expected to occur~\cite{Ueda_Review_2018,Schmiedmayer_Review_2016}. The dynamical critical point in the prethermal regime is also relevant for experiments in the near future, as the thermalization timescale is likely to be far beyond the experimental coherence time~\cite{neyenhuis2017,kinoshita2006,gring2012}.

\begin{figure}
\includegraphics[width=\linewidth]{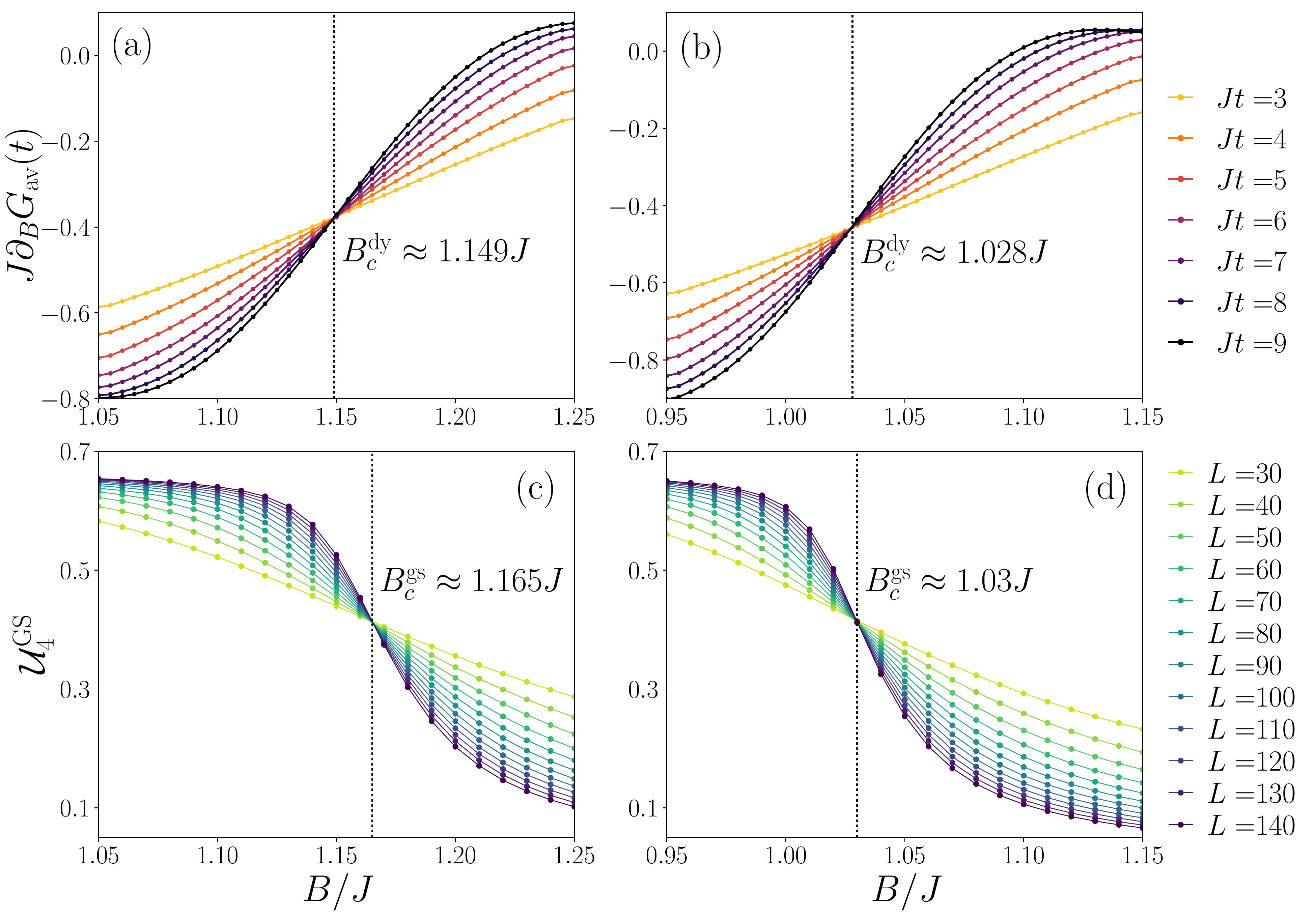}
\caption{Comparison between the finite-time scaling of the time-averaged nearest-neighbor correlator $G_{\rm av}(t)$ and the ground-state Binder cumulant $\mathcal{U}_4^{\rm gs}$. The first column [panels (a) and (c)] corresponds to the TFIM with next-nearest-neighbor interaction $\Delta=0.1$ [\cref{eq:model-NNN}], while the second column [panels (b) and (d)] corresponds to the TFIM with long-range interaction $\alpha=6$ [\cref{eq:model-LR}]. The finite-time scaling of the quench dynamics is shown in the first row, and the finite-size scaling for the ground state is shown in the second row. The dynamical critical points ($B_c^{\rm dy}$) identified in (a) and (b) using finite-time scaling are close but different from the locations of the quantum critical points ($B_c^{\rm gs}$) identified in (c) and (d). The ground-state simulations were done using a DMRG algorithm with bond dimension $\chi=32$.}
\label{fig:prethermalization_derivative}
\end{figure}

We now show that our finite-time scaling method can also reveal a dynamical critical point in the prethermal regime of a nearly integrable system. We calculate the non-integrable dynamics numerically due to the lack of analytic solutions. Using a split-operator decomposition (using fourth-order Suzuki-Trotter expansion) coupled with a Walsh-Hadamard transform, we calculate dynamics for up to $L=25$ spins (see~\cite{supp} for technical details). In Figs.~\ref{fig:prethermalization_derivative}(a) and (b) we plot the time dependence of the derivative of $G_{\rm av}(t)$ for $\Delta=0.1$ and $\alpha=6$ respectively. Remarkably, the curves at different times still cross at one point, which we identify as the dynamical critical point $B_c^{\rm dy}$. These curves will also collapse almost perfectly on top of each other using a scaling function $\partial_B G_{\rm av}(t)=\frac{1}{J}\tilde{f}\left((B-B_c^{\rm dy})t\right)$ \cite{supp}, similar to the integrable case.

In the integrable TFIM, we showed that the dynamical critical point $B_c^{\rm dy}$ coincides with the ground-state critical point $B_c^{\rm gs}$. It is natural to ask whether this is still the case for the above-calculated models near integrability. To locate the quantum critical point, we compute the Binder cumulant~\cite{Binder1981}, $\mathcal{U}^{\rm gs}_{4}=1-\frac{\langle M^4\rangle_{\rm gs}}{3\langle M^2\rangle^2_{\rm gs}}$ (where $M=\frac{1}{L}\sum_i\sigma^x_i$ and $\langle\cdots\rangle_{\rm gs}$ denotes the ground-state expectation value) using a DMRG algorithm for a system with open boundary conditions and sizes ranging from $L=30$ to $L=140$. We identify the quantum critical point using a finite-size scaling method as shown in Figs.~\ref{fig:prethermalization_derivative}(c) and (d). It is found that $B_c^{\rm dy}$ is close, but not identical, to $B_c^{\rm gs}$ at both $\Delta=0.1$ and $\alpha=6$.

While we cannot make a conclusive statement about whether $B_c^{\rm dy}$ agrees with $B_c^{\rm gs}$ in the thermodynamic limit based on finite-size numerical calculations, we argue that $B_c^{\rm dy}$ and $B_c^{\rm gs}$ should in general be different but close to each other when near integrability. To support this argument, we perform a self-consistent mean-field calculation \cite{Sen_Chakrabarti_1991} for \cref{eq:model-NNN} with $\Delta=0.1$ to identify both $B_c^{\rm dy}$ and $B_c^{\rm gs}$ in the thermodynamic limit \cite{supp}. The next-nearest-neighbor spin interaction translates to a perturbative two-particle interaction of the Jordan-Wigner fermions. The essence of the self-consistent calculation is to approximate this interaction by  effective single-particle hoppings and on-site energies. This makes the Hamiltonian noninteracting, with the quench dynamics given by an effective GGE\@. The parameters of this effective Hamiltonian must be determined self-consistently from the expectation values of different correlation functions. These expectation values may be considered either in the ground state or the effective GGE corresponding to the quench from some initial state. While the former determines the quantum critical point  $B_c^{\rm gs}$~\cite{Sen_Chakrabarti_1991}, we claim that the latter captures the dynamical critical point $B_c^{\rm dy}$. Therefore, it is natural to expect a difference in the locations of the dynamical and quantum critical points.

The self-consistent mean-field calculation should be asymptotically exact as $\Delta\rightarrow 0$. We find that to first order in $\Delta$, $B^{\rm dy}_c\approx J\left(1+\frac{3}{2}\Delta\right)$ and $B^{\rm gs}_c\approx J\left(1+\frac{16}{3\pi}\Delta\right)$ \cite{supp}. For $\Delta=0.1$, these predict $B_c^{\rm dy}\approx 1.15$ and $B_c^{\rm gs}\approx 1.168$, agreeing very well with numerics in  Figs.~\ref{fig:prethermalization_derivative}(a) and (c). We have further confirmed the accuracy of these predictions for $\Delta$ up to 0.3 \cite{supp}. For the larger values of $\Delta$, it is clear that $B^{\rm dy}_c\neq B^{\rm gs}_c$, but they are close in magnitude when $\Delta \approx 0$. 

{\it Discussion.}---%
Our results are relevant to experiments in trapped ions~\cite{Zhang_Monroe_2017} and Rydberg atoms~\cite{Lukin_Rydberg_2017}, where dynamics of the TFIM can be readily probed. As an example, we numerically obtain the dynamics for the $\alpha=6$ TFIM, which models Rydberg atoms interacting via van der Waals type long-range interactions~\cite{Rydberg_Review_2010}. We find that the numerically obtained dynamical critical point $B^{\rm dy}_c\approx 1.028J$ shown in \cref{fig:prethermalization_derivative}(b) is very close to the ground-state critical point $B^{\rm gs}_c\approx 1.03J$ obtained from finite-size scaling shown in \cref{fig:prethermalization_derivative}(d). We emphasize that the dynamical critical point is identified by our finite-time scaling method for a system size of only 25 spins and an evolution time of only $9/J$, which are well within the current experimental record for the system size and coherence time \cite{Lukin_Rydberg_2017}. As a result, we believe our method can be used in near-future experiments to identify critical points of quantum phase transitions using quench dynamics.

Our work opens up several interesting questions for future consideration: (i) How should one classify the observed dynamical critical point? We note that the dynamical critical point identified using $G(t)$ as an ``order parameter'' does not represent a conventional symmetry-breaking phase transition, because for both $B<B_c^{\rm dy}$ and $B>B_c^{\rm dy}$, the quenched state does not spontaneously break the Ising symmetry and become ferromagnetically ordered. (ii) Is finite-time scaling a general method for identifying dynamical critical points? We believe that for generic, short-range interacting systems, finite-time scaling serves the purpose of finite-size scaling for finding the quantum critical point due to the emergence of linear light cones \cite{lieb1972}. However, when interactions become long-range, the linear light cone may no longer exist \cite{fossfeig2015,eldredge2017} and it remains unclear when the finite-time scaling method fails. (iii) Could the link between dynamical and quantum critical points established here be generalized to other integrable and nearly integrable systems, such as systems solvable by Bethe ansatz or many-body localized systems? Here the link is provided by the single-particle spectrum that governs both equilibrium and nonequilibrium physics, but what happens when the single-particle spectrum is less relevant? (iv) Could there be similar links between dynamical phase transitions and phase transitions in excited eigenstates? This is particularly relevant in many-body localized systems \cite{Nandksihore_Huse_2013}. (v) Can the prethermal dynamical critical point be predicted using other theoretical methods such as the kinetic equations using time-dependent GGEs~\cite{Alessio_Review_2016,Bertini_Robinson_2015} or Keldysh field theory?

\begin{acknowledgments}
{\it Acknowledgments.}---%
We thank John Bollinger, Lincoln Carr, Fabian Essler, Mohammad Maghrebi, Toma\v{z} Prosen, Marcos Rigol, Somendra M. Bhattacharjee, Sthitadhi Roy, and Soonwon Choi for useful discussions.  The authors have been financially supported by the NSF Ideas Lab on Quantum Computing, ARO, AFOSR, NSF PFC at JQI, ARO MURI, ARL CDQI, the DOE ASCR Quantum Testbed Pathfinder program, and the DOE BES Materials and Chemical Sciences Research for Quantum Information Science program. P.T. and J.R.G. acknowledge support from the NIST NRC Research Postdoctoral Associateship Award. Z.X.G. acknowledges the start-up fund support from Colorado School of Mines. This research was supported in part by the National Science Foundation under Grant No. NSF PHY-1748958. This work used the Extreme Science and Engineering Discovery Environment (XSEDE), which is supported by National Science Foundation grant number ACI-1548562.  Specifically, it used the Bridges system, which is supported by NSF award number ACI-1445606, at the Pittsburgh Supercomputing Center (PSC)~\cite{xsede, bridges}. The authors acknowledge the University of Maryland supercomputing resources (http://hpcc.umd.edu) made available for conducting the research reported in this Letter.
\end{acknowledgments}

\bibliographystyle{apsrev4-1}
\bibliography{dptrefs} 



\section*{Supplementary material}

Paraj Titum, Joseph T. Iosue, James R. Garrison, Alexey V. Gorshkov, and Zhe-Xuan Gong

In this Supplemental Material, we provide details of calculations referenced in the main text. Section S.I focuses on analytical results of the nearest-neighbor transverse-field Ising model (TFIM), deriving Eqs. (4) and (6) and Fig. 2 of the main text and discussing the finite-size and finite-time scaling properties of the derivative of the nearest-neighbor correlator. In Sec. S.II, we provide details of the self-consistent mean-field calculations, which were used in the main text to estimate the location of the critical point in the TFIM with next-nearest-neighbor interactions. In Sec. S.III, we remark briefly on the scaling properties in the presence of integrability breaking to explain the scaling collapse in Fig. 3 of the main text. In Sec. S.IV, we describe the numerical method we used for time evolution.

## S.I. ANALYTICAL CALCULATION DETAILS FOR NEAREST-NEIGHBOR TFIM

In this section, we provide details for the analytical results of the nearest-neighbor transverse field Ising model. In Secs. S.I A and S.I B, we review the well known method of diagonalizing the Hamiltonian and representing the initial state. In Sec. S.I C, we calculate the analytical formula for the time dependence of the nearest-neighbor correlator, $G(t)$. Finally, in Sec. S.I D, we compare the finite-time and finite-size scaling of the derivative, $\frac{\partial G_{\text{av}}(t)}{\partial B}$.

### A. Diagonalizing the Hamiltonian

Under the Jordan-Wigner transformation, the nearest-neighbor transverse-field Ising model with periodic boundary conditions [Eq. (1) with $\Delta = 0$] transforms to a free fermion model,

$$H_\eta = J \sum_{i=1}^{L-1} \left( a_i - a_i^\dagger \right) \left( a_{i+1} + a_{i+1}^\dagger \right)$$
$$+ B \sum_{i=1}^{L} (2a_i^\dagger a_i - 1) + \eta \left( a_L - a_L^\dagger \right) \left( a_1 + a_1^\dagger \right). \quad (S1)$$

Here, $\eta = \exp\left(i\pi \sum_j a_j^\dagger a_j\right)$ is the parity of the fermion number, with $\eta = 1$ for an even number of fermions and $\eta = -1$ for an odd number of fermions. We can now take advantage of the translation symmetry in $H_\eta$ by using a Fourier transformation, $b_{q_\eta} = \frac{1}{\sqrt{L}} \sum_{j=1}^{L} a_j e^{iq_\eta j}$ where $q_\eta = \frac{\pi}{L}\left[2m + \frac{1}{2}(\eta - 1)\right]$ with $m \in \{-\frac{L}{2}+1, \cdots 0, 1, \cdots, \frac{L}{2}\}$.

This gives us

$$H_\eta = 2J \sum_{q_\eta > 0} \mathbf{b}^\dagger \begin{pmatrix} \frac{B}{J} - \cos q_\eta & i \sin q_\eta \\ -i \sin q_\eta & -\frac{B}{J} + \cos q_\eta \end{pmatrix} \mathbf{b}, \quad (S2)$$

with $\mathbf{b} = \begin{pmatrix} b_{q_\eta} & b_{-q_\eta}^\dagger \end{pmatrix}^T$. This can now be diagonalized using a Bogoliubov transformation, $\gamma_{q_\eta} = u_{q_\eta} b_{q_\eta} + i v_{q_\eta} b_{-q_\eta}^\dagger$, with $u_{q_\eta} = \cos \theta_{q_\eta}$, $v_{q_\eta} = \sin \theta_{q_\eta}$, and $\tan 2\theta_{q_\eta} = \frac{J \sin q_\eta}{B - J \cos q_\eta}$. In this basis we obtain the diagonal Hamiltonian

$$H_\eta = \sum_{q_\eta > 0} \omega_{q_\eta} \left( \gamma_{q_\eta}^\dagger \gamma_{q_\eta} + \gamma_{-q_\eta}^\dagger \gamma_{-q_\eta} - 1 \right), \quad (S3)$$

with $\omega_{q_\eta} = 2J\sqrt{\frac{B^2}{J^2} - 2\frac{B}{J}\cos q_\eta + 1}$. The ground state is $|\phi_0\rangle \equiv \frac{1}{|v_{q_\eta}|} \prod_{q_\eta > 0} \gamma_{q_\eta} \gamma_{-q_\eta} |0\rangle$, where $|0\rangle \equiv |\downarrow \cdots \downarrow\rangle$ is the vacuum state of the $a_i$ operators. All excited states can be written down as a set of excitations with momenta denoted by $\{Q_\eta\}$:

$$|\phi_n\rangle = \prod_{q_\eta \in \{Q_\eta\}} \gamma_{q_\eta}^\dagger |\phi_0\rangle = \prod_{q_\eta} |p_{n,q_\eta} p_{n,-q_\eta}\rangle, \quad (S4)$$

where $p_{n,q_\eta} = \langle \phi_n | \hat{I}_{q_\eta} | \phi_n \rangle = 0$ or $1$, denoting the occupation of the mode with momentum $q_\eta$.

### B. Initial state and expectation values

We choose the initial state to have all spins polarized along the $x$ direction:

$$|\psi_{\text{in}}\rangle = |\rightarrow_x \cdots \rightarrow_x\rangle = \frac{1}{2^{L/2}} \left(1 + a_1^\dagger\right) \cdots \left(1 + a_L^\dagger\right) |0\rangle. \quad (S5)$$

Note that the initial state $|\psi_{\text{in}}\rangle$ is a superposition of both even ($\eta = 1$) and odd ($\eta = -1$) parity sectors. We can write the initial state as an equal-weighted superposition of both parity sectors, $|\psi_{\text{in}}\rangle = \frac{1}{\sqrt{2}} \left(|\psi_{\text{in}}^{\eta=1}\rangle + |\psi_{\text{in}}^{\eta=-1}\rangle\right)$. We provide analytical expressions for the expectation values of various oper-

ators in this initial state projected to a given parity sector:

$$\left\langle b_{q_\eta}^\dagger b_{q_\eta} \right\rangle_{\text{in},\eta} = \frac{1}{2L} \left[ L + (L-2) \cos q_\eta \right], \quad \text{(S6)}$$

$$\left\langle b_{-q_\eta} b_{-q_\eta}^\dagger \right\rangle_{\text{in},\eta} = \frac{1}{2L} \left[ L - (L-2) \cos q_\eta \right], \quad \text{(S7)}$$

$$\left\langle b_{q_\eta}^\dagger b_{-q_\eta}^\dagger \right\rangle_{\text{in},\eta} = \frac{i(L-2)}{2L} \sin q_\eta, \quad \text{(S8)}$$

$$\left\langle b_{-q_\eta} b_{q_\eta} \right\rangle_{\text{in},\eta} = -\frac{i(L-2)}{2L} \sin q_\eta, \quad \text{(S9)}$$

$$\left\langle \gamma_{q_\eta}^\dagger \gamma_{q_\eta} \right\rangle_{\text{in},\eta} = \frac{1}{2L} \left[ L - 2(L-2) \frac{J - B \cos q_\eta}{\omega_{q_\eta}} \right], \quad \text{(S10)}$$

$$\left\langle \gamma_{q_\eta} \gamma_{q_\eta} \right\rangle_{\text{in},\eta} = -\frac{i(L-2) \sin q_\eta}{L \omega_{q_\eta}} B, \quad \text{(S11)}$$

where $\langle \cdots \rangle_{\text{in},\eta} = \langle \psi_{\text{in}}^\eta | \cdots | \psi_{\text{in}}^\eta \rangle$, and $|\psi_{\text{in}}^\eta\rangle$ is assumed to be normalized.

### C. Time dependence of $G(t)$

In this section, we obtain the time dependence of the correlator analytically. The nearest-neighbor correlator was defined in Eq. (3) in the main text. Using the Jordan-Wigner transformation, we obtain an expression for the operator that measures the nearest-neighbor correlator,

$$\hat{G}_\eta = -\frac{1}{L} \sum_{q_\eta > 0} (-2 \cos q_\eta) \left( b_{q_\eta}^\dagger b_{q_\eta} - b_{-q_\eta} b_{-q_\eta}^\dagger \right) + 2i \sin q_\eta \left( b_{q_\eta} b_{-q_\eta} + b_{q_\eta}^\dagger b_{-q_\eta}^\dagger \right). \quad \text{(S12)}$$

Note that this expression depends upon the fermion number parity $\eta$ through the definition of $q_\eta$. The correlation operator $\hat{G}_\eta$ does not change the parity of the state it acts on. Therefore the expectation value of the time-dependent correlator can be written as

$$G(t) = \frac{1}{2} \left[ \langle \psi_{\text{in}}^{\eta=1} | \hat{G}_1(t) | \psi_{\text{in}}^{\eta=1} \rangle + \langle \psi_{\text{in}}^{\eta=-1} | \hat{G}_{-1}(t) | \psi_{\text{in}}^{\eta=-1} \rangle \right]. \quad \text{(S13)}$$

Recall from the previous section that the expectation values of the conserved densities, $I_{q_\eta}$, in the initial state are only dependent on the parity through $q_\eta$, just like in Eq. (S12). Therefore, the analytical expression for the contribution from both sectors will be the same and we can drop the dependence on $\eta$ in $q_\eta$ in all expressions and evaluate the formula of $G(t)$ using Eqs. (S6) to (S11). Going to the Bogoliubov basis, we can exactly calculate the time dependence,

$$\hat{G}(t) = \hat{G}_0 + \frac{1}{L} \sum_{q>0} \frac{4B \sin q}{\omega_q} i \left( \gamma_{-q} \gamma_q e^{-2i\omega_q t} + \text{H.c.} \right), \quad \text{(S14)}$$

where the $q$ sum is taken over values allowed for either parity, and where

$$\hat{G}_0 = \sum_q \frac{1 - 2\gamma_q^\dagger \gamma_q}{\omega_q} \left[ -2 \cos q (B - J \cos q) + 2J \sin^2 q \right]. \quad \text{(S15)}$$

Taking an expectation value with respect to the initial state recovers the analytical expressions for $G_{\text{av}}(t)$ and $G_{\text{av,arbitrary}}^\infty$ in Eqs. (4) and (6) of the main text. Now, we can obtain an analytical expression for the first derivative of $G_{\text{av}}(t)$ with respect to $B$:

$$\frac{\partial G_{\text{av}}(t)}{\partial B} = \frac{(L-2)}{L^2} \sum_q \frac{16B \sin^2 q}{\omega_q^4} \Big[ 2J(B \cos q - J) + \\ + B(B - J \cos q) \cos(2t\omega_q) \\ - (B^2 + BJ \cos q - 2J^2) \frac{\sin(2t\omega_q)}{2t\omega_q} \Big]. \quad \text{(S16)}$$

In the main text, we use Eq. (S16) to obtain the time dependence of the first derivative in Fig. 2.

### D. Scaling of the derivative, $\frac{\partial G_{\text{av}}(t)}{\partial B}$

#### 1. Finite-time scaling

Let us consider the finite-time scaling of the first derivative of the correlator defined in Eq. (S16). We are interested in the behavior near the critical point. With this in mind, let us consider an expansion around $B = J + \epsilon$. Noting that $\frac{\partial G_{\text{av}}(t)}{\partial B}\Big|_{B=J} = -\frac{1}{2J}$, we expand around the critical point,

$$\frac{\partial G_{\text{av}}(t)}{\partial B}\bigg|_{B=J+\epsilon} \approx -\frac{1}{2J} \int_0^\pi \frac{dq}{2\pi} \left( (1+\cos q) \left[ 2 - \cos(2\omega_q t) - \frac{\sin(2\omega_q t)}{2\omega_q t} \right] \\ -\frac{\epsilon}{J} \cot^2\left(\frac{q}{2}\right) \left[ 2 + \left(1 + \sin^2\left(\frac{q}{2}\right)\right) \cos(2\omega_q t) - \left(64J^2 t^2 \sin^4\left(\frac{q}{2}\right) + \sin^2\left(\frac{q}{2}\right) + 3\right) \frac{\sin(2\omega_q t)}{2\omega_q t} \right] \right), \quad \text{(S17)}$$

where we have used $\omega_q = 2J\sqrt{2(1-\cos q)}$ and taken the thermodynamic limit. Let us examine the above expansion term by term. At $O(\epsilon^0)$, the integrand contains terms $\propto$

$\cos(2\omega_q t)$ and $\frac{\sin(\omega_q t)}{2\omega_q t}$, which average out to zero as time increases. This explains why the derivative is very weakly dependent on time at the critical point ($\epsilon = 0$). Going to the next order at $O(\epsilon)$, we need to be careful around $q \to 0$, since the summand has terms $\propto \frac{1}{\sin^2(q/2)} \sim \frac{1}{q^2}$. Expanding this term around $q = 0$, we obtain the following expression for the derivative near the critical point,

$$\frac{\partial G_{\text{av}}(t)}{\partial B} \approx -\frac{1}{2J} + \frac{\epsilon t}{J}\mathcal{F}(Jt), \qquad \text{(S18)}$$

with

$$\mathcal{F}(x) = \frac{1}{x}\int_0^\pi \frac{dq}{2\pi}\frac{2}{q^2}\left(2 + \cos(4xq) - 3\frac{\sin(4xq)}{4xq}\right). \qquad \text{(S19)}$$

Note that in the limit of large $x$, $\mathcal{F}(x)$ approaches 1, and we can expand $\frac{\partial G_{\text{av}}(t)}{\partial B}$ near $B = J$ as

$$\left.\frac{\partial G_{\text{av}}(t)}{\partial B}\right|_{B=J+\epsilon} = \frac{1}{J}\left(-\frac{1}{2} + \epsilon t\right). \qquad \text{(S20)}$$

This linear dependence is shown as a red dashed line in the inset of Fig. 2 in the main text.

### 2. Finite-size scaling

Now let us discuss the comparison between finite-time scaling and finite-size scaling. In order to isolate the scaling with system size, we examine the infinite-time-averaged value of the first derivative, $\frac{\partial G_{\text{av}}(t\to\infty)}{\partial B}$. Utilizing the expression for the derivative expanded around the critical point (but for finite system sizes), we obtain the following expression for the infinite-time-averaged value of the derivative:

$$\left.\frac{\partial G_{\text{av}}(t\to\infty)}{\partial B}\right|_{B=J+\epsilon} = -\frac{1}{2J} + \frac{\epsilon}{J^2 L}\sum_{\eta,q_\eta>0}\frac{4}{q_\eta^2}$$
$$\approx \frac{1}{J}\left(-\frac{1}{2} + \frac{\epsilon}{3J}L\right). \qquad \text{(S21)}$$

Note that for the above expression, the different parity sectors ($\eta = \pm 1$) contribute differently to the coefficient of $\epsilon$. Compared to Eq. (S20), we find that the time $t$ plays the role of $L/(3J)$, where $v = 3J$ can be regarded as a Lieb-Robinson velocity. As a result, the finite-time scaling is in some sense equivalent to the finite-size scaling.

## S.II. SELF-CONSISTENT MEAN-FIELD CALCULATIONS FOR TFIM WITH NEXT-NEAREST-NEIGHBOR INTERACTIONS

In this section, we provide details of the self-consistent mean-field calculations we performed to estimate the critical point of the TFIM in the presence of next-nearest-neighbor interactions. In general, this model cannot be mapped to a free-fermion model, and therefore we need to make some approximations to calculate its critical point analytically.

Before doing a self-consistent calculation, let us outline properties of a simpler, noninteracting Hamiltonian,

$$\tilde{H} = J\sum_i\left(a_i - a_i^\dagger\right)\left(a_{i+1} + a_{i+1}^\dagger\right) + J\Delta\sum_i\left(a_i - a_i^\dagger\right)\left(a_{i+2} + a_{i+2}^\dagger\right) + B\sum_{i=1}^L\left(2a_i^\dagger a_i - 1\right). \qquad \text{(S22)}$$

Equation (S22) can be diagonalized in a straightforward manner after Fourier transformation to obtain the quasiparticle spectrum

$$\omega_q = 2J\sqrt{g^2 + 1 + \Delta^2 - 2g(\cos q + \Delta\cos 2q) + 2\Delta\cos q}, \qquad \text{(S23)}$$

where we have defined $g \equiv B/J$. The quasiparticle annihilation operator is defined analogously as $\gamma_q = \cos\theta_{q_\eta}b_q + i\sin\theta_{q_\eta}b_{-q}^\dagger$, with $\tan 2\theta_{q_\eta} = \frac{\sin q + \Delta\sin 2q}{g - \cos q - \Delta\cos 2q}$ and $b_q = \frac{1}{\sqrt{L}}\sum_j a_j e^{iqj}$. The critical point is at $g = 1 + \Delta$, where it is easy to verify that $\omega_q = 0$ has a solution at $q = 0$. We can now calculate long-time ($t \to \infty$) values of different correlators for dynamics under the Hamiltonian given by Eq. (S22):

$$\mathcal{O}_0 = -\frac{1}{L}\sum_i\left\langle 2a_i^\dagger a_i - 1\right\rangle_{\text{in},t\to\infty} = \frac{2J}{L}\sum_q\frac{1-2\langle I_q\rangle_{\text{in}}}{\omega_q}(g - \cos q - \Delta\cos 2q), \qquad \text{(S24)}$$

$$\mathcal{O}_1 = -\frac{1}{L}\sum_i\left\langle\left(a_i - a_i^\dagger\right)\left(a_{i+1} + a_{i+1}^\dagger\right)\right\rangle_{\text{in},t\to\infty} = \frac{2J}{L}\sum_q\frac{1-2\langle I_q\rangle_{\text{in}}}{\omega_q}[(\Delta - g)\cos q + 1], \qquad \text{(S25)}$$

$$\mathcal{O}_2 = -\frac{1}{L}\sum_i\left\langle\left(a_i - a_i^\dagger\right)\left(a_{i+2} + a_{i+2}^\dagger\right)\right\rangle_{\text{in},t\to\infty} = \frac{2J}{L}\sum_q\frac{1-2\langle I_q\rangle_{\text{in}}}{\omega_q}[(\Delta - g\cos 2q + \cos q)], \qquad \text{(S26)}$$

where $\langle \cdots \rangle_\text{in}$ indicates expectation value with respect to the initial state, and $I_q = \gamma_q^\dagger \gamma_q$ is the quasiparticle density at momentum $q$. Examining the functional dependence of the correlators, it is clear that starting from the initial state $|\rightarrow \cdots \rightarrow\rangle$, the long-time values of the operators will exhibit a nonanalyticity at the critical point given by $B/J = 1 + \Delta$.

Now let us discuss the Ising model with next-nearest-neighbor interactions as defined in Eq. (1). Under the Jordan-Wigner transformation, it becomes

$$H_\text{NNN} = J \sum_i \left(a_i - a_i^\dagger\right)\left(a_{i+1} + a_{i+1}^\dagger\right) + J\Delta \sum_i \left(a_i - a_i^\dagger\right)\left(1 - 2a_{i+1}^\dagger a_{i+1}\right)\left(a_{i+2} + a_{i+2}^\dagger\right) + B \sum_{i=1}^L \left(2a_i^\dagger a_i - 1\right). \quad \text{(S27)}$$

In order to do a self-consistent calculation, we approximate the four-fermion term with a 'Hartree-Fock' like expansion [S1]. While this is standard when considering the ground state of $H_\text{NNN}$, we extend this approximation to the long-time quench dynamics by taking the mean-field expectation values in the $t \to \infty$ limit with respect to the initial state $|\psi_\text{in}\rangle$:

$$\begin{aligned}
\left(a_i - a_i^\dagger\right)\left(1 - 2a_{i+1}^\dagger a_{i+1}\right)\left(a_{i+2} + a_{i+2}^\dagger\right) &\approx \left\langle \left(a_i - a_i^\dagger\right)\left(a_{i+2} + a_{i+2}^\dagger\right)\right\rangle_{\text{in},t\to\infty} \left(1 - 2a_{i+1}^\dagger a_{i+1}\right) \\
&+ \left\langle \left(1 - 2a_{i+1}^\dagger a_{i+1}\right)\right\rangle_{\text{in},t\to\infty} \left(a_i - a_i^\dagger\right)\left(a_{i+2} + a_{i+2}^\dagger\right) \\
&- \left\langle \left(a_i - a_i^\dagger\right)\left(a_{i+1} + a_{i+1}^\dagger\right)\right\rangle_{\text{in},t\to\infty} \left(a_{i+1} - a_{i+1}^\dagger\right)\left(a_{i+2} + a_{i+2}^\dagger\right) \\
&- \left(a_i - a_i^\dagger\right)\left(a_{i+1} + a_{i+1}^\dagger\right) \left\langle \left(a_{i+1} - a_{i+1}^\dagger\right)\left(a_{i+2} + a_{i+2}^\dagger\right)\right\rangle_{\text{in},t\to\infty}.
\end{aligned} \quad \text{(S28)}$$

Plugging this expansion back into the Hamiltonian, we obtain an effective Hamiltonian of the form given by $\tilde{H}$ in Eq. (S22) but with the following the modified parameters:

$$J' = J + 2J\Delta \mathcal{O}'_1, \quad \text{(S29)}$$
$$J'\Delta' = J\Delta \mathcal{O}'_0, \quad \text{(S30)}$$
$$B' = B + J\Delta \mathcal{O}'_2, \quad \text{(S31)}$$

where $\mathcal{O}'_i$ are computed using Eqs. (S24) to (S26) with the replacement $\{J, \Delta, B\} \to \{J', \Delta', B'\}$. The critical point is given by $\Delta' = 1 + g'$, which coupled with the above equations gives us the value of the field $B$ at the critical point. In the following, we will discuss two possible cases of solving these self-consistent equations:

1. **Ground state**: For the ground state of Eq. (S27) after the mean-field approximation, we have $\langle I_q\rangle_\text{gs} = 0$. We obtain the following system of equations for the critical point:

$$J' = J(1 + 4\Delta I_1), \quad \text{(S32)}$$
$$2J\Delta I_1 = (g' - 1)(J(1 + 4\Delta I_1) - 2J\Delta I_2), \quad \text{(S33)}$$
$$B = g'J' - 2J\Delta(g'I_2 - I_1), \quad \text{(S34)}$$

with $I_1 = J' \int_{-\pi}^\pi \frac{dq}{2\pi} \frac{1}{\omega'_q}(-\cos q + 1)$ and $I_2 = J' \int_{-\pi}^\pi \frac{dq}{2\pi} \frac{1}{\omega'_q}(-\cos 2q + 1)$, and where $\omega'_q$ is given by Eq. (S23) with $\{J, \Delta, B\} \to \{J', \Delta', B'\}$. When $\Delta$ is small, these self-consistent equations can be solved analytically to obtain $B \approx J + \frac{16}{3\pi}J\Delta$. For general $\Delta$, these equations can be solved numerically. For instance, at $\Delta = 0.1$, we obtain $B/J = 1.16$ as the critical point.

2. **Quench from** $|\rightarrow \cdots \rightarrow\rangle$: In this case, $\langle I_q\rangle_\text{in} = \frac{1}{2} - \frac{J'}{\omega'_q}[1 - (g' - \Delta')\cos q]$. Again, the critical point is

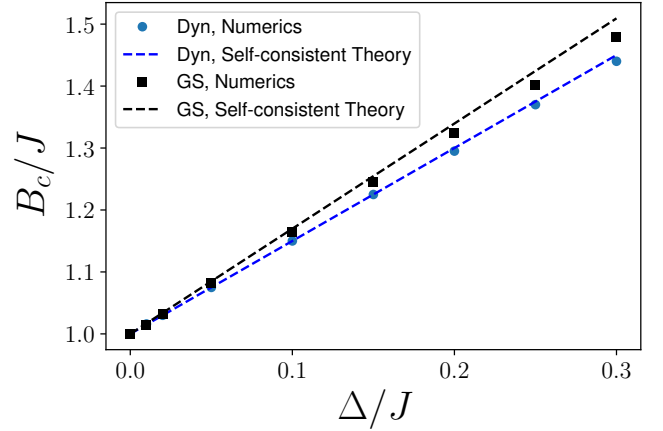

FIG. S1. Comparison between the self-consistent theory and exact numerics. Blue points show the dynamical critical points obtained from finite-time scaling as a function of the integrability breaking. In comparison, the blue dashed line shows the prediction from the mean-field approximation. Black points show the quantum critical point obtained from the finite-size scaling of the Binder cumulant, $\mathcal{U}_4$. The black dashed line shows the expected dependence from the self-consistent mean-field approximation for the ground state.

given by

$$(g' - 1)(g' + \Delta) = \Delta \frac{2g' - 1}{2g'}, \quad \text{(S35)}$$
$$B = J(g' + \Delta). \quad \text{(S36)}$$

The above equations can be solved to first order in $\Delta$, which gives us the critical point at $B \approx J + \frac{3}{2}J\Delta$.

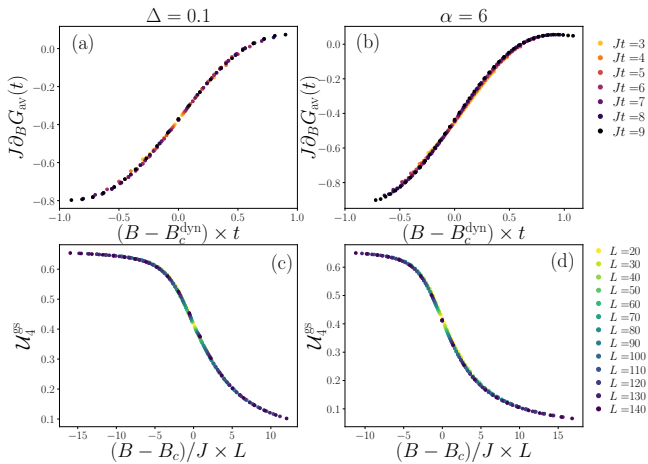

FIG. S2. Scaling collapse with parameters given by the respective panels of Fig. 3 in the main text. The first column [panels (a) and (c)] corresponds to the TFIM with next-nearest-neighbor interaction $\Delta = 0.1$ [Eq. (1)], while the second column [panels (b) and (d)] corresponds to the TFIM with long-range interaction $\alpha = 6$ [Eq. (2)]. The top panels [(a) and (b)] show finite-time scaling of the time-averaged nearest-neighbor correlator $G_{\mathrm{av}}(t)$, while the bottom panels [(c) and (d)] show finite-size scaling of the ground-state Binder cumulant $\mathcal{U}_4^{\mathrm{gs}}$.

Note that in general, the self-consistent equations give different solutions for the critical point when using the ground-state or the quench dynamics. In Fig. S1, we compare the predictions for the critical points from the self-consistent theory (both ground-state and quench) with the actual critical points obtained from numerics. We locate the critical points for different values of $\Delta$ using the technique employed for Figs. 3(a) and (c) of the main text. Figure S1 shows reasonable agreement between the self-consistent theory and the numerical values.

## S.III. SCALING COLLAPSE FOR FIG. 3

In this section, we briefly remark on the scaling properties in the presence of integrability breaking. In Fig. S2, we show the scaling collapse under rescaling of the $x$ axis of Fig. 3. The important thing to note is the distinction between Fig. S2 panels (a) and (b) in comparison with (c) and (d). While the dynamical data collapses to a universal function around the dynamical critical point with $\frac{\partial G_{\mathrm{av}}(t)}{\partial B} = \tilde{f}\left((B - B_c^{\mathrm{dy}})t\right)$, the ground-state Binder cumulant collapses to a different universal function around the quantum critical point, $\mathcal{U}_4^{\mathrm{gs}} = K((B - B_c^{\mathrm{gs}})L/J)$.

## S.IV. NUMERICAL ALGORITHM FOR TIME EVOLUTION

In this section, we present the numerical method we used for performing time evolution. We assume a spin-half model on $N$ sites in which the Hamiltonian can be decomposed as

$$H = H_x + H_z, \quad \text{(S37)}$$

with $H_x$ diagonal in the $x$ basis and $H_z$ diagonal in the $z$ basis. The method takes advantage of the identity $X = \mathsf{H}Z\mathsf{H}$, where $\mathsf{H} = \frac{1}{\sqrt{2}}\begin{pmatrix} 1 & 1 \\ 1 & -1 \end{pmatrix}$ is the Hadamard operator, and $X$ and $Z$ are Pauli matrices. It relies on a fast Walsh-Hadamard transform to apply $\mathsf{H}$ to all spins simultaneously. A similar method has been discussed for simulating a kicked Ising model [S2].

Let us first consider the method using a first-order Trotter decomposition,

$$e^{-iH\tau} = e^{-iH_x\tau}e^{-iH_z\tau} + O(\tau^2). \quad \text{(S38)}$$

We now consider implementing this operator, which approximates evolution for a time $\tau$, on a state vector $|\psi\rangle$ which is represented in the $z$ basis. The operator $e^{-iH_z\tau}$ can be applied in $O(2^N)$ time since it is diagonal in the $z$ basis. It remains to apply $e^{-iH_x\tau}$. Note that the operator can be written in the form

$$e^{-iH_x\tau} = \mathsf{H}_{\mathrm{all}} e^{-iD\tau} \mathsf{H}_{\mathrm{all}}, \quad \text{(S39)}$$

with $D = \mathsf{H}_{\mathrm{all}} H_x \mathsf{H}_{\mathrm{all}}$ diagonal in the $z$ basis, and where $\mathsf{H}_{\mathrm{all}} = \prod_j \mathsf{H}_j$ is the Hadamard operator applied to all sites $j$. The operator $\mathsf{H}_{\mathrm{all}}$ can be applied to a state in time $O(2^N N)$ and without memory overhead via the fast Walsh-Hadamard transform, which is faster than the equivalent dense matrix multiplication. In this way, the full Trotter time step can be applied by performing two element-wise multiplications of the state vector to implement the diagonal operators, along with two Walsh-Hadamard transforms. The Walsh-Hadamard transform can even be performed in parallel across many cores or nodes in a cluster, e.g. as implemented in FFTW [S3]. If the diagonal elements of $H_x$ and $H_z$ are computed on the fly, the method results in zero memory overhead, unlike Krylov-space methods.

In fact, we can improve the error due to the finite time step by using a fourth-order Suzuki-Trotter decomposition [S4],

$$e^{-iH\tau} = U(\tau_1)U(\tau_1)U(\tau_3)U(\tau_1)U(\tau_1) + O(\tau^5), \quad \text{(S40)}$$

where $U(\tau_i) = e^{-iH_z\tau_i/2}e^{-iH_x\tau_i}e^{-iH_z\tau_i/2}$, $\tau_1 = \frac{\tau}{4-4^{1/3}}$, and $\tau_3 = \tau - 4\tau_1$. (Note that $\tau_3$ is actually negative.) The simulations within this paper were obtained by using such a fourth-order decomposition with $\tau = 0.005$. Benchmarking against exact diagonalization suggests that this time step is sufficiently small for the simulations in this paper.

We note that given a maximum permissible error of $\epsilon$ (as quantified by the trace distance), the number of Walsh-Hadamard transforms required to simulate for time $T$ is at most $O(T^{5/4}\epsilon^{1/4})$ for such a fourth-order decomposition [S5]. (Here we assume that error from the Suzuki-Trotter decomposition dominates and error from the Walsh-Hadamard transform is negligible.) Recent evidence suggests that this rigorous error bound may not be tight (see, e.g., Ref. [S6]).

Finally, we note that generalizations of the Walsh-Hadamard transform can be used to implement additional terms in the Hamiltonian, e.g., a term $H_y$ that is diagonal in the $y$ basis.



\end{document}